\documentclass[journal,trans]{IEEEtran}
\usepackage{float}
\usepackage{soul}
\usepackage{mathtools}
\usepackage{epsfig,wrapfig}
\usepackage{epstopdf} 
\usepackage{caption}
\usepackage{subcaption}
\usepackage{fancyhdr}
\usepackage{psfrag}
\usepackage{lipsum}
\usepackage{xcolor}
\usepackage{multirow}
\usepackage{booktabs}
\usepackage{rotfloat}
\usepackage{cite}
\usepackage{amsmath}
\usepackage{cite}
\usepackage{cuted}

\definecolor{cadmiumgreen}{rgb}{0.0, 0.42, 0.24}
\DeclareUnicodeCharacter{2212}{-}

\newcommand{\mt}{\mathrm}

\renewcommand{\parallel}{\mathrel{/\mkern-5mu/}}
\makeatletter
\newcommand{\notparallel}{%
  \mathrel{\mathpalette\not@parallel\relax}%
}
\newcommand{\pati}[1]{\reversemarginpar\hspace{0pt}\marginpar{\hspace{-2.2cm}\begin{minipage}{2.8cm}\centering\vspace{\baselineskip}\textcolor{magenta}{\textit{\phantom{#1}}}\end{minipage}}}
\ifCLASSINFOpdf
\else
\fi
\begin{document}

\title{Space-Time Wedges}

\author{\IEEEauthorblockN{Amir Bahrami,
Klaas De Kinder,
Zhiyu Li and Christophe Caloz,~\IEEEmembership{Fellow,~IEEE}}

}

\IEEEtitleabstractindextext{%
\begin{abstract}
Space-time-modulated systems have attracted significant interest over the past decade due to their ability to manipulate electromagnetic waves in unprecedented ways. Here, we introduce a new type of space-time-modulated structure, the \emph{space-time wedge}, consisting of two interfaces moving at different velocities, which results in either closing or opening wedges. Using moving boundary conditions, we derive closed-form solutions for the scattering of electromagnetic waves in such a wedge and leverage these solutions to unveil the underlying physics, including multiple space-time scattering and Doppler shifting. The space-time wedge holds potential for various optical and photonic applications.
\end{abstract}

}
\maketitle
\IEEEdisplaynontitleabstractindextext
\IEEEpeerreviewmaketitle


\section{Introduction}\label{sec:intro}
 
\pati{GSTEMs}
Generalized Space-Time Engineered Modulation (GSTEM) systems, or GSTEMs for short, are structures whose properties are modulated in both space and time by an external drive~\cite{caloz2022gstem}. The modulation can take various forms, including electronic, optical, acoustic, mechanical, thermal and chemical~\cite{rhodes1981acousto,saleh2019fundamentals,xu2022diffusive,Shaltout_Science_2019,tessier2023experimental}. This modulation typically manifests as a traveling or standing wave perturbation in one of the medium's constitutive parameters. Therefore, GSTEM systems are best classified based on their modulation velocity regime. The most common regime is instantaneous modulation (or infinite velocity)~\cite{Morgenthaler_1958,mirmoosa2024temporal}, which enables a wide range of applications and physical phenomena, such as time reversal~\cite{fink2016timereversal,mirmoosa2024temporal}, breaking of fundamental bounds~\cite{shlivinski2018beyond}, beam splitting~\cite{mendoncca2003temporal}, photon generation~\cite{mendoncca2000quantum} and cooling~\cite{pendry2024air}, inverse prism~\cite{akbarzadeh2018inverse}, parametric amplification~\cite{tien1958parametric}, temporal impedance matching~\cite{engheta2020antireflection} and temporal aiming~\cite{engheta2020aiming}. In recent years, this regime has also been explored beyond classical physics~\cite{hilbert2009temporal,reck2017dirac,goldman2014periodically,dong2024quantum,ok2024electron}. The modulation velocity can also vary uniformly, ranging from subluminal to superluminal speeds~\cite{bolotovskiui1972superlum,deck2019uniform,pendry2022crossing}, which introduces additional novel phenomena, including Doppler shifting~\cite{granatstein1976realization,lampe1978interaction,deck2019uniform}, magnetless nonreciprocity~\cite{hadad2024space,taravati2017nonreciprocal,estep2014magnetic}, space-time reversal~\cite{Deck_PRB_2018}, dynamic diffraction~\cite{taravati2019generalized}, asymmetric bandgaps~\cite{cassedy1963dispersion,cassedy1967dispersion,deck2019uniform} and isolation~\cite{chamanara2017optical}, light deflection~\cite{huidobro2019fresnel,huidobro2021homogenization,Lurie_Springer_2007} and shock-wave production~\cite{joannopoulos2003shockwave}. Finally, the modulation velocity can be nonuniform, where acceleration enables phenomena such as photon emission~\cite{joannopoulus2022accelerating}, chirping~\cite{bahrami2023generalized}, light bending~\cite{bahrami2023bending} and gravity analogs~\cite{bahrami2023electrodynamics}.

\pati{Basic GSTEM Structures}
GSTEMs encompass several fundamental structures, including interfaces, slabs, space-time crystals and space-time metamaterials. Interfaces serve as the core building blocks of all GSTEMs~\cite{caloz2019spacetime1,caloz2019spacetime2}. Slabs are formed by stacking two interfaces moving at the same velocity~\cite{bellman1966slab,tsai1967wave}. Space-time crystals result from the periodic repetition of slabs with different properties~\cite{deck2019uniform}. Finally, space-time metamaterials are created by reducing the spatial and temporal periods of these crystals to subwavelength and subperiod scales~\cite{cassedy1963dispersion,deck2019uniform}.

\pati{Key Contribution}
Here, we introduce a new fundamental class of GSTEM structures, the \emph{space-time wedge}. A space-time wedge is formed by combining two space-time interfaces with different velocities, corresponding to a wedge- or triangular-shaped structure in the space-time diagram. In a purely spatial representation, with space as the abscissa and a property (such as refractive index or potential) as the ordinate, these wedges correspond to shrinking (closing wedge) or expanding (opening wedge) slabs.

\pati{Organization}
The paper is organized as follows. Section~\ref{sec:concept} introduces the concept of space-time wedges as an extension of conventional space-space wedges. Then, Sec.~\ref{sec:classification} presents a classification of all possible types of space-time wedges. Next, Sec.~\ref{sec:methodology} provides the resolution strategy to determine the scattered waves at a space-time wedge. Then, Sec.~\ref{sec:scattering} addresses the scattering problem at dielectric space-time wedges and provides two representative examples for such phenomena. Next, Sec.~\ref{sec:frequency} describes the frequency transitions observed in the scattered waves. Section~\ref{sec:impenetrable} tackles the scattering problem at impenetrable wedges as a special case of dielectric wedges, with examples showcasing space-time multiple scattering effects and Doppler frequency shifts. Finally, Sec.~\ref{sec:discussion} discusses experimental implementations and potential applications.

\section{Space-Time Wedge Concept}\label{sec:concept}
 
\begin{figure}[!h]
    \centering
    \vspace{-4mm}
    \includegraphics[width=0.5\textwidth]{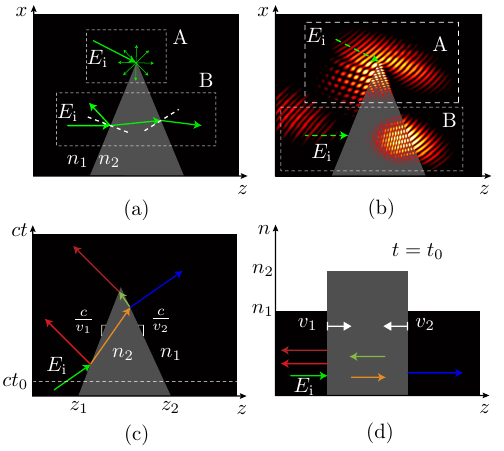}\vspace{-4mm}
    \caption{Wedge structure, consisting of two interfaces separating two media of different refractive indices, $n_1$ and $n_2$, and related scattering. (a)~Conventional space-space closing wedge with (b)~diffraction due to vertex excitation and refraction due to edge excitation. (c)~Space-time wedge, with two interfaces moving at different velocities, $v_1$ and $v_2$, and colors corresponding to different frequencies, and (d)~related space-index perspective.}\label{fig:Structure}
\end{figure}

\pati{Wedge Concept Introduction}
The space-time wedge must be clearly distinguished from the conventional, ``space-space'' wedge, in terms of both its structure and its operation. Let us therefore describe both wedges, with the help of Fig.~\ref{fig:Structure}.

\pati{Space-Space Wedges}
Figure~\ref{fig:Structure}(a) shows the geometry of a space-space wedge. Such a wedge is a dielectric or metallic structure with one or more sharp vertices~\cite{ishimaru2017electromagnetic}. When light impinges on a vertex, it diffracts, while when it impinges on an edge, it undergoes Snell's refraction, as shown in Figs.~\ref{fig:Structure}(a) and~\ref{fig:Structure}(b). The optical behavior of space-space wedges, including diffraction and refraction phenomena, has been extensively studied and is considered a canonical problem in electromagnetic theory~\cite{saleh2019fundamentals}.

\pati{Space-Time Wedges}
Figure~\ref{fig:Structure}(c) shows a space-time wedge. This wedge is obtained by replacing the space ($x$) ordinate of the space-space wedge in Fig.~\ref{fig:Structure}(a) by time ($ct$). The resulting structure consists of two interfaces moving at different velocities, whose space-index representation is shown in Fig.~\ref{fig:Structure}(d)\footnote{In quantum physics, this structure would be interpreted as a widening or narrowing potential barrier or well.}. Unlike the space-space wedge, the space-time wedge does not involve diffraction, due to causality. In contrast, it induces Doppler shifting\footnote{The reflection Doppler shifting is well-known, with up/down-shifting for an approaching/receding object. The transmission Doppler shifting is less known; it depends not only on the velocity of the object's interface, but also on its refractive index contrast. This may be understood in terms of the pulse expansion and compression described in~\cite{caloz2019spacetime1}.} in both reflection and transmission, which can be either upshifted or downshifted depending on the velocity and properties of the medium.

\pati{Assumption}
In this paper, we restrict ourselves to space-time wedges with \emph{two edges} (and one vertex), originating at the spatial points $z_1$ and $z_2$, as shown in Fig.~\ref{fig:Structure}(d). However, an analogous approach may be applied to the problem of wedges with two or three vertices, and even to the problem of polygonal structures with more edges.
 
\section{Classification}\label{sec:classification}
 
\pati{Interface Classification}
Space-time wedges, as described in Sec.~\ref{sec:concept}, are composed of two space-time interfaces. These interfaces can be classified according to their velocity, $v$. The velocity may be subluminal, interluminal, or superluminal. The subluminal and superluminal velocities correspond to regimes $v<\mt{min}\{c/n_1,c/n_2\}$ and $v>\mt{max}\{c/n_1,c/n_2\}$, respectively, while the interluminal velocities  correspond to the regime $\mt{min}\{c/n_1,c/n_2\}<v<\mt{max}\{c/n_2,c/n_2\}$. The scattering behavior of interfaces in the subluminal and superluminal regimes have been extensively studied~\cite{Lurie_Springer_2007,caloz2019spacetime2}. In contrast, investigations of scattering in the interluminal regime remain limited to specific cases~\cite{ostrovskiicorrect1967,deck2019interluminal}.

\begin{figure}[!h]
    \centering
    \vspace{-4mm}
    \includegraphics[width=0.5\textwidth]{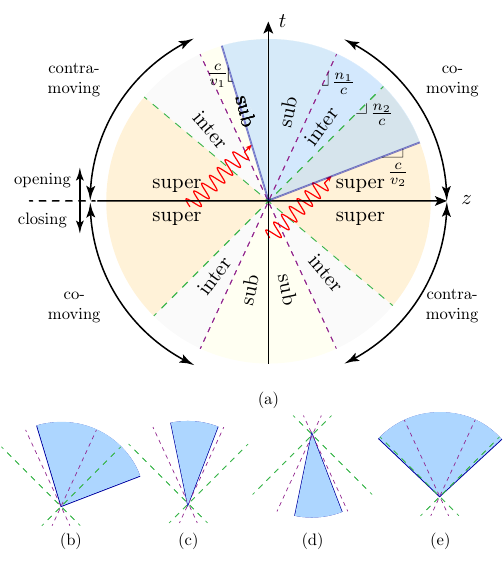}\vspace{-3mm}
    \caption{Classification of space-time wedges. (a)~Generic operating regimes and examples of (b)~contra-sub co-sup opening, (c)~contra-sub co-sub opening, (d)~co-sub contra-sub closing and (d)~contra-sup co-sup opening wedges.}\label{fig:Bigpicture}
\end{figure}

\pati{Space-Time Wedge Classification}
The wedges can be classified according to the velocity regimes (subluminal, interluminal, or superluminal) of their two interfaces and whether the wedge is opening or closing with time. Figure~\ref{fig:Bigpicture} represents all the possible types of wedges together with the assumed direction of the incident wave ($\pm{c}/n_1$). For instance, the wedge represented in Fig.~\ref{fig:Bigpicture}(a), and isolated in Fig.~\ref{fig:Bigpicture}(b), corresponds to an opening wedge, whose left interface is subluminal and contra-moving with respect to the incident wave, and right interface is superluminal and comoving with respect to the incident wave. Such a configuration will be hereafter referred to as a contra-sub co-sup opening wedge. Similarly, Figs.~\ref{fig:Bigpicture}(c),~(d) and~(e) correspond to contra-sub co-sub opening, co-sub contra-sub closing and contra-sup co-sup opening wedges, respectively. This classification results in 78 distinct scattering configurations (see Sec.~1 in~\cite{supp_mat} for all the configurations).

\section{Resolution Frame Selection}\label{sec:methodology}
 
\pati{Inapplicability of Frame-Hopping}
Scattering problems at moving interfaces are commonly addressed using a method known as frame-hopping~\cite{van2012relativity}. This approach simplifies the problem by transforming coordinates from the laboratory frame to the comoving frame, reducing it to the scattering at a stationary interface. Once the scattering coefficients are calculated in this comoving frame, the results are then transformed back to the laboratory frame. This technique is particularly effective for analyzing scenarios involving moving interfaces and slabs~\cite{deck2019uniform}. However, in the case of wedges, we encounter a complication: wedges entail two velocities that are generally not equal\footnote{In the limiting case of equal velocities, the wedge reduces to a slab.}. Consequently, there is no frame where both interfaces are stationary\footnote{This situation is akin to two cars moving in opposite directions, making it impossible to follow both simultaneously.}. As a result, we must tackle the problem within the laboratory frame and use moving boundary conditions.
 
\section{Scattering Formulas}\label{sec:scattering}
 
\pati{Waveform Assumptions}
According to Fig.~\ref{fig:Structure}(c), the incident ($\psi_\mt{i}$), transmitted ($\psi_\mt{t}$) and reflected ($\psi_\mt{r}$) waves in medium~1 (surrounding of the wedge) are traveling waves, which may be written as
\begin{subequations}\label{eq:wave_ass}
    \begin{equation}\label{eq:incident}
        E^\mt{i}_x=\psi_\mt{i}\left[\frac{z}{u_1}-t\right],
    \end{equation}
    \begin{equation}\label{eq:transmitted}
        E^\mt{t}_x=\psi_\mt{t}\left[\frac{z}{u_1}-t\right]
    \end{equation}
and
    \begin{equation}\label{eq:reflected}
        E^\mt{r}_x=\psi_\mt{r}\left[\frac{z}{u_1}+t\right],
    \end{equation}
\end{subequations}
where $u_1=c/n_1$ is the speed of light in medium $1$, where we have assumed that the incident wave is launched in medium~1. On the other hand, waves in medium~2 (wedge) are a superposition of forward ($\psi_\mt{f}$) and backward ($\psi_\mt{b}$) traveling waves, which may be written as
\begin{equation}\label{eq:wedge_ass}
    E^\mt{w}_x=\psi_\mt{f}\left[\frac{z}{u_2}-t\right]+\psi_\mt{b}\left[\frac{z}{u_2}+t\right],
\end{equation}
where $u_2=c/n_2$ is the speed of light in medium $2$. We have used square brackets, $\left[\cdot\right]$, to indicate the arguments of the wave functions ($\psi_\mt{i}$, $\psi_\mt{t}$, $\psi_\mt{r}$, $\psi_\mt{f}$ and $\psi_\mt{b}$) in Eqs.~\eqref{eq:wave_ass} and~\eqref{eq:wedge_ass}. In the forthcoming examples, we will use Gaussian modulated pulse for the incidence, viz., $\smash{\psi_\mt{i}[z/u_1-t]=\mt{exp}({\mt{i}\omega_\mt{i}(z/u_1-t)})\mt{exp}({-\frac{(z/u_1-t)^2}{\sigma^2}})}$, with $\omega_\mt{i}$ being the incident frequency and $\sigma$ the width of the Gaussian pulse.

\pati{Main Derivations}
We now apply the moving boundary conditions at each interface of the wedge. These conditions prescribe the continuity of $\mathbf{E}+\mathbf{v_1}\times\mathbf{B}$ and $\mathbf{H}-\mathbf{v_1}\times\mathbf{D}$ for the first interface and $\mathbf{E}+\mathbf{v_2}\times\mathbf{B}$ and $\mathbf{H}-\mathbf{v_2}\times\mathbf{D}$ for the second interface. This leads to a pair of recurrent equations whose combination provides the sought-after scattered wave solutions (see Sec.~2 in~\cite{supp_mat} for detailed derivations)
\begin{subequations}\label{eq:scatt}
\begin{align}\label{eq:scatt_trans}
\begin{split}
    &\psi_\mt{t}\left[\frac{z}{u_1}-t\right]=\\&T_{12}T_{21}H\sum_{p=0}^{\infty}R^{2p}D^p\psi_\mt{i}\underbrace{\left[\Delta\phi_p+HD^p\left(\frac{z}{u_1}-t\right)\right]}_{\Phi_\mt{t}}
\end{split}
\end{align}
and
\begin{align}\label{eq:scatt_refl}
\begin{split}
&\psi_\mt{r}\left[\frac{z}{u_1}+t\right]= R\frac{M^-_{11}}{M^+_{11}}\psi_\mt{i}\left[\phi'_0-\frac{M^-_{11}}{M^+_{11}}\left(\frac{z}{u_1}+t\right)\right]
\\& -T_{12}T_{21}H'\sum_{p=0}^{\infty}R^{2p+1}D^{'p}\psi_\mt{i}\underbrace{\left[\Delta\phi'_p+H'D^{'p}\left(\frac{z}{u_1}+t\right)\right]}_{\Phi_\mt{r}}
\end{split}
\end{align}
with
\begin{equation}\label{eq:scatt_phase_t}
    \Delta\phi_p=A+BC\left(\frac{1-D^p}{1-D}\right)+BD^p\left(\frac{z_2}{u_2}-\frac{z_2M^-_{22}}{u_1M^-_{21}}\right),
\end{equation}
\begin{equation}
    \phi'_0=\frac{z_1}{u_1}\left(1+\frac{M^-_{11}}{M^+_{11}}\right)
\end{equation}
and
\begin{equation}\label{eq:scatt_phase_r}
    \Delta\phi'_p=A'+B'C'\left(\frac{1-D^{'p}}{1-D'}\right)+B'D^{'p }\left(\frac{z_1}{u_2}-\frac{z_1M^+_{12}}{u_1M^+_{11}}\right),
\end{equation}
where
\begin{equation}
    R=\frac{\eta_2-\eta_1}{\eta_2+\eta_1 },\quad T_{ij}=\frac{2\eta_j}{\eta_i+\eta_j},\quad M^\pm_{ij}=1\pm\frac{v_i}{u_j},
\end{equation}
\begin{equation}\label{eq:aux_a}
    A=\frac{z_1}{u_1}-\frac{z_1}{u_2}B,\quad A'=\frac{z_1}{u_1}-\frac{z_2}{u_2}B'+\frac{z_2-z_1}{u_2}\frac{M^-_{11}}{M^-_{12}},
\end{equation}
\begin{equation}\label{eq:aux_b}
    B=\frac{M^-_{11}}{M^-_{12}},\quad B'=-\frac{M^-_{11}M^-_{22}}{M^-_{12}M^+_{22}},
\end{equation}
\begin{equation}\label{eq:aux_c}
    C=\frac{z_2}{u_2}-\frac{z_1}{u_2}D+\frac{z_2-z_1}{u_2}\frac{M^-_{22}}{M^+_{22}},
\end{equation}
\begin{equation}\label{eq:aux_cp_d}
    C'=\frac{z_1}{u_2}-\frac{z_2}{u_2}D'-\frac{z_2-z_1}{u_2}\frac{M^+_{12}}{M^-_{12}},\quad D=D'=\frac{M^+_{12}M^-_{22}}{M^-_{12}M^+_{22}}
\end{equation}
and
\begin{equation}\label{eq:aux_h}
    H=\frac{M^-_{22}M^-_{11}}{M^-_{21}M^-_{12}},\quad H'=-\frac{M^+_{12}M^-_{11}M^-_{22}}{M^+_{11}M^-_{12}M^+_{22}}.
\end{equation}
\end{subequations}
In these relations, $R$ and $T_{ij}$ correspond to the conventional reflection and transmission coefficients at a stationary interface (from media $1$ to $2$, resp. $i$ to $j$) with $\eta_1$ and $\eta_2$ being the impedance of medium~1 and~2, $M_{ij}$ is a term related to Doppler shifting, and $\Delta\phi_p$ and $\Delta\phi'_p$ together with the parameters given in Eqs.~\eqref{eq:aux_a}-\eqref{eq:aux_h} account for the accumulated phase shifts for the transmitted and the reflected waves, respectively.

\pati{Multiple Space-Time Scattering}
Equations~\eqref{eq:scatt} may be physically interpreted as follows. The summation terms in Eqs.~\eqref{eq:scatt_trans} and~\eqref{eq:scatt_refl}, running from $0$ to $\infty$, represent the \emph{multiple space-time scattering events} occurring within the wedge [see Fig.~\ref{fig:Structure}~(d)]. Let us first analyze the reflected wave, Eq.~\eqref{eq:scatt_refl}. The first, summation-less term in the right-hand side of Eq.~\eqref{eq:scatt_refl}, which has no equivalent in Eq.~\eqref{eq:scatt_trans}, corresponds to the initial reflection at the first interface, while the second term, with the factor $T_{12}T_{21}H'$ describes transmission across the wedge, with the effect of the multiple scattering within the wedge being accounted for by the summation including reflection $R^{2p+1}$ and Doppler shifting $D^{'p}$. Similar observations can be made with the transmitted wave, Eq.~\eqref{eq:scatt_trans}, since the corresponding relation is essentially similar to the reflected wave.

\pati{Dielectric Example}
Figure~\ref{fig:Dielectric} shows the space-time evolution and frequency spectra of the scattering phenomena described by Eqs.~\eqref{eq:scatt_trans} and~\eqref{eq:scatt_refl}. Figure~\ref{fig:Dielectric}(a) shows the scattering behavior of a closing wedge. Upon encountering the initial interface, a portion of the incident wave is reflected ($E_{\mt{r},1}$) with a Doppler downshift, attributed to the interaction with a comoving interface. Within the wedge, the wave undergoes multiple contramoving space-time reflections, resulting in successive Doppler upshifting. In contrast, Fig.~\ref{fig:Dielectric}(b) presents the scattering behavior of an opening wedge. Here, the initial reflection ($E_{\mt{r},1}$) exhibits upshifting. As the wave propagates within the wedge, it experiences successive Doppler downshifting due to repeated interactions with the comoving interfaces.

\begin{strip}
    \centering
    \includegraphics[width=1\textwidth]{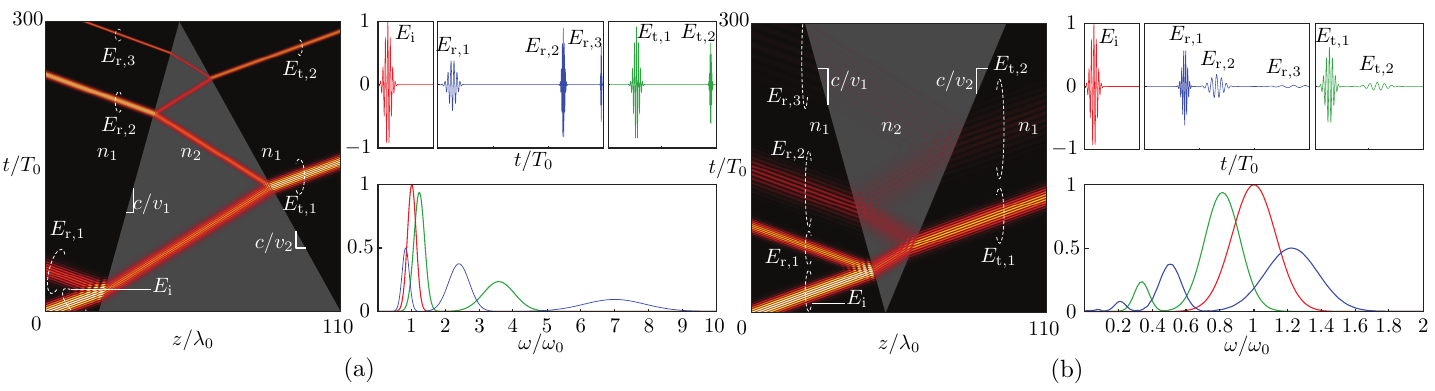}\vspace{-3mm}
    \captionof{figure}{Scattering at a dielectric wedge separating two media, of permittivities $\epsilon_1=1$ and $\epsilon_2=3$. (a)~Closing wedge, with $v_1=0.1c$ and $v_2=-0.2c$, and (b)~opening wedge, with $v_1=-0.1c$ and $v_2=0.15c$.}\label{fig:Dielectric}
\end{strip}
 
\section{Frequency Transitions}\label{sec:frequency}
 
\pati{Frequency Analysis of the Scattered Waves}
Space-time wedges produce scattered waves with multiple new frequencies, as illustrated in Fig.~\ref{fig:Dielectric}. The related frequency shifts result from multiple Doppler shifts occurring during interactions with the wedge interfaces. The frequency change after each scattering event can be calculated by taking the derivative of the arguments in Eqs.~\eqref{eq:scatt_trans} and~\eqref{eq:scatt_refl}, viz.,

\begin{subequations}\label{eq:Freq}
    \begin{equation}\label{eq:Freq_Trans}
        f_\mt{t}=\frac{\partial\Phi_\mt{t}}{\partial t}=HD^p
    \end{equation}
and
    \begin{equation}\label{eq:Freq_Refl}
        f_\mt{r}=\frac{\partial\Phi_\mt{r}}{\partial t}=H'D^{'p},
    \end{equation}
\end{subequations}
where $p$ denotes the number of scattering events. After each interaction, the frequency of the wave is either upshifted or downshifted, depending on whether it interacts with a contramoving or a comoving interface, respectively.

\begin{figure}[!ht]
    \centering
    \includegraphics[width=0.4\textwidth]{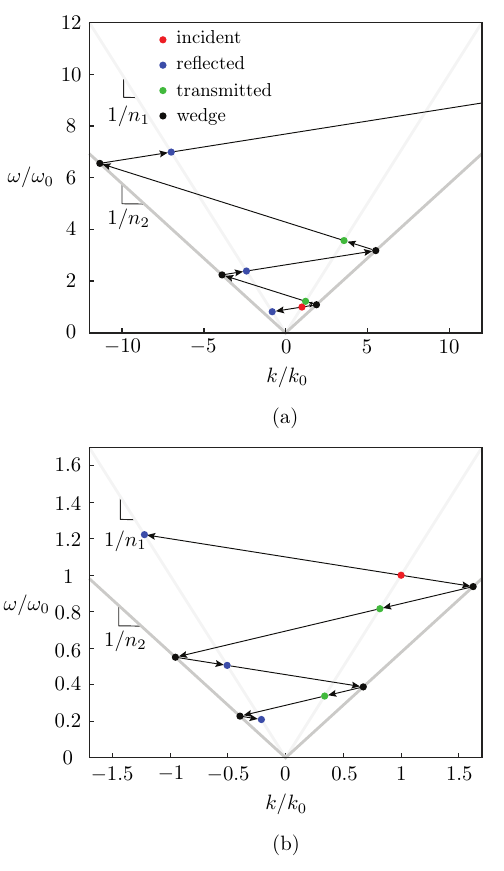}\vspace{-3mm}
    \caption{Transition diagrams for the cases of (a)~a closing wedge with velocities $v_1=0.1c$ and $v_2=-0.2c$ [Fig.~\ref{fig:Dielectric}(a)] and (b)~a closing wedge with velocities~$v_1=-0.1c$ and $v_2=0.15c$ [Fig.~\ref{fig:Dielectric}(b)].}\label{fig:Freq_Tran}
\end{figure}

\pati{Graphical Approach}
These frequency transformations [Eq.~\eqref{eq:Freq}] can be graphically represented and validated in the ``\emph{transition diagrams}'' shown in Fig.~\ref{fig:Freq_Tran}, where Figs.~\ref{fig:Freq_Tran}(a) and~\ref{fig:Freq_Tran}(b) corresponds to the closing and opening wedges of Figs.~\ref{fig:Dielectric}(a) and~\ref{fig:Dielectric}(b). The transition diagrams of Figs.~\ref{fig:Freq_Tran} may be progressively constructed by following the wave scattering events in Figs.~\ref{fig:Dielectric}. Let us consider the closing wedge case. The incident wave starts at the point $(k/k_0,\omega/\omega_0)=(1,1)$ (red dot). Then, it experiences a first pair of transitions at the interface~1, and hence under the angle $v_1$: a reflection transition in medium $n_1$ (lowest blue dot) and a transmission transition into medium $n_2$ (lowest black dot). Next, the transmitted wave (lowest black dot) crosses the wedge and incurs a new scattering event and a new transition pair at the interface~2, under the angle $v_2$: a transmission transition into medium $n_1$ (lowest green dot) and a reflection transition in medium $n_2$ (second lowest black dot). The next scattering points, and Fig.~\ref{fig:Freq_Tran}(b), follow the same logic. Note that the closing wedge gradually increases the reflected and transmitted frequencies--or photon energies--while the opening wedge constantly decreases these frequencies or energies. 

\section{Impenetrable Wedges}\label{sec:impenetrable}
 
\pati{PEC Wedge}
Having examined wave scattering and propagation in space-time wedges formed by modulated dielectric materials, we now focus on a particular case of space-time wedge, a space-time wedge with its exterior (medium~1) being impenetrable\footnote{A space-time wedge with its \emph{interior} (medium~2) being impenetrable is possible but fairly trivial, as its scattering reduces to a single (Doppler-shifted) reflection.}. In contrast to dielectric wedges, where waves propagate and scatter \emph{across} the space-time structure, the impenetrable space-time wedge entirely confines waves within its interior once excited there. The scattering and propagation of electromagnetic waves in the structure can be found by applying \emph{Perfect Electric Conductor (PEC) moving boundary conditions} at each interface~\cite{deck2019uniform}, viz., $\mathbf{E}+\mathbf{v_1}\times\mathbf{B}=0$ at the first interface and $\mathbf{E}+\mathbf{v_2}\times\mathbf{B}=0$ at the second interface. This operation results in a recursive equation, whose solution is the multiple scattering expression\footnote{Note that Eq.~\eqref{eq:p_scatt} is \emph{not} a particular case of Eqs.~\eqref{eq:scatt_trans} and~\eqref{eq:scatt_refl}. This is because reflection in the PEC wedge occurs exclusively within the wedge structure whereas the dielectric wedge equations apply only outside the wedge.}~\cite{supp_mat}
\begin{align}\label{eq:p_scatt}
\begin{split}
E^\mt{w}_x=&\sum_{p=0}^{\infty} (R_1R_2)^{p+1} \psi_\mt{i}\left[\Delta \phi_p +(R_1R_2)^{p+1}\left(\frac{z}{u_1}-t\right)\right]\\&-\sum_{p=0}^{\infty} R_1^pR_2^{p+1} \psi_\mt{i}\left[\Delta\phi'_p-R_1^pR_2^{p+1}\left(\frac{z}{u_1}+t\right)\right],
\end{split}
\end{align}
where
\begin{subequations}
    \begin{equation}
        R_1=\frac{M^+_{11}}{M^-_{11}}
    \end{equation}
    and
    \begin{equation}
        R_2=\frac{M^-_{21}}{M^+_{21}}
    \end{equation}
\end{subequations}
are the conventional scattering coefficients at a PEC interface moving at velocity $v_1$ and $v_2$, respectively~\cite{tsai1967wave}.

\pati{Illustration of PEC Wedge}
Figure~\ref{fig:PEC} shows the space-time evolution and frequency spectra of scattering at space-time wedges with impenetrable (PEC) interfaces, computed by Eq.~\eqref{eq:p_scatt} and validated by FDTD simulation~\cite{bahrami2023generalized}. Unlike the dielectric wedges discussed in Sec.~\ref{sec:scattering}, where the wave may escape the wedge, the wave in the PEC wedges remain confined within the structure. Figure~\ref{fig:PEC}(a) depicts the scattering in a closing PEC wedge, where the wave constantly interacts with contramoving interfaces, causing a gradual frequency upshift. Conversely, Fig.~\ref{fig:PEC}(b) shows scattering in an opening PEC wedge, where the wave gradually reflects off comoving interfaces, resulting in a gradual frequency downshift.

\begin{strip}
  \centering
  \includegraphics[width=\linewidth]{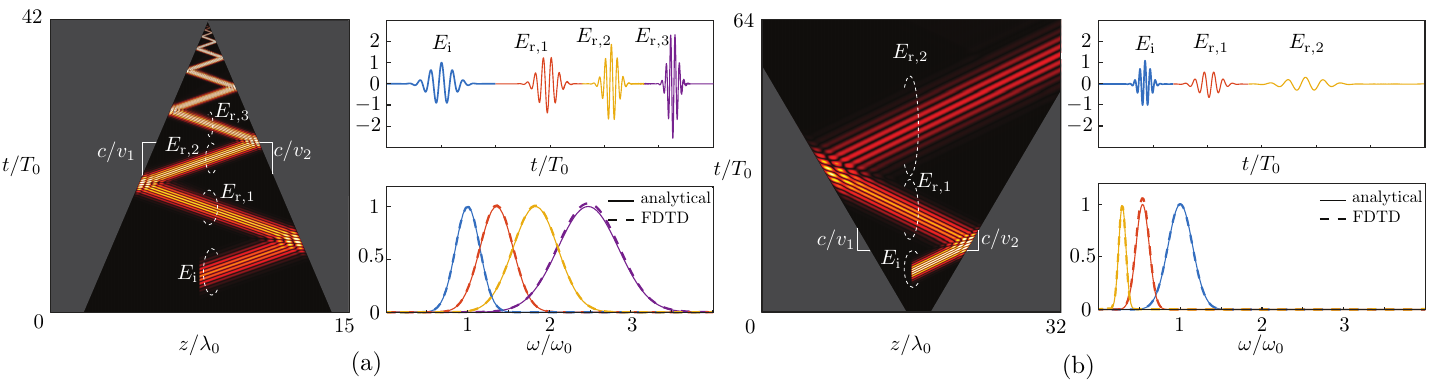} 
  \captionof{figure}{Scattering at PEC space-time wedges shown in space-time diagrams (left panels) with corresponding time-domain waveforms and spectra (right panels) for (a)~a closing wedge, with interfaces moving at velocities $v_1=-v_2=0.15c$ and (b)~an opening wedge with interfaces moving at velocities $v_1=-v_2=-0.3c$.}
  \label{fig:PEC}
\end{strip}

\section{Conclusion}\label{sec:discussion}
 
\pati{Conclusion}
The paper has introduced and explored the concept of space-time wedges, a novel type of GSTEM. We have demonstrated the unique scattering phenomena that occur in these structures, including multiple Doppler frequency shifts and multiple space-scattering scattering. We have classified the various types of space-time wedges based on their velocity regimes and derived closed-form solutions for the scattered waveforms. Such a wedge may have various applications in dynamic classical optical and quantum photonic devices, such as frequency converters, modulators, multiplexers, wave or particle traps, and photon heating and cooling systems.

\ifCLASSOPTIONcaptionsoff
  \newpage
\fi


{\small
\bibliographystyle{IEEEtran}
\bibliography{ref_wedge.bib}

\providecommand{\noopsort}[1]{}\providecommand{\singleletter}[1]{#1}%
\begin{thebibliography}{10}
\providecommand{\url}[1]{#1}
\csname url@samestyle\endcsname
\providecommand{\newblock}{\relax}
\providecommand{\bibinfo}[2]{#2}
\providecommand{\BIBentrySTDinterwordspacing}{\spaceskip=0pt\relax}
\providecommand{\BIBentryALTinterwordstretchfactor}{4}
\providecommand{\BIBentryALTinterwordspacing}{\spaceskip=\fontdimen2\font plus
\BIBentryALTinterwordstretchfactor\fontdimen3\font minus \fontdimen4\font\relax}
\providecommand{\BIBforeignlanguage}[2]{{%
\expandafter\ifx\csname l@#1\endcsname\relax
\typeout{** WARNING: IEEEtran.bst: No hyphenation pattern has been}%
\typeout{** loaded for the language `#1'. Using the pattern for}%
\typeout{** the default language instead.}%
\else
\language=\csname l@#1\endcsname
\fi
#2}}
\providecommand{\BIBdecl}{\relax}
\BIBdecl

\bibitem{caloz2022gstem}
C.~Caloz, Z.-L. Deck-L{\'e}ger, A.~Bahrami, O.~C. Vicente, and Z.~Li, ``Generalized space-time engineered modulation ({GSTEM}) metamaterials: {A} global and extended perspective.'' \emph{IEEE Antennas Propag. Mag.}, vol.~65, no.~4, pp. 50--60, 2023.

\bibitem{rhodes1981acousto}
W.~T. Rhodes, ``Acousto-optic signal processing: {C}onvolution and correlation,'' \emph{Proc. IEEE}, vol.~69, no.~1, pp. 65--79, 1981.

\bibitem{saleh2019fundamentals}
B.~E. Saleh and M.~C. Teich, \emph{Fundamentals of {P}hotonics}.\hskip 1em plus 0.5em minus 0.4em\relax John Wiley \& Sons, 2019.

\bibitem{xu2022diffusive}
L.~Xu, G.~Xu, J.~Huang, and C.-W. Qiu, ``Diffusive {F}izeau drag in spatiotemporal thermal metamaterials,'' \emph{Phys. Rev. Lett.}, vol. 128, no.~14, p. 145901, 2022.

\bibitem{Shaltout_Science_2019}
A.~M. Shaltout, V.~M. Shalaev, and M.~L. Brongersma, ``Spatiotemporal light control with active metasurfaces,'' \emph{Science}, vol. 364, no. 6441, pp. 1--10, May 2019.

\bibitem{tessier2023experimental}
S.~Tessier~Brothelande, C.~Cro{\"e}nne, F.~Allein, J.~O. Vasseur, M.~Amberg, F.~Giraud, and B.~Dubus, ``Experimental evidence of nonreciprocal propagation in space-time modulated piezoelectric phononic crystals,'' \emph{Appl. Phys. Lett.}, vol. 123, no.~20, 2023.

\bibitem{Morgenthaler_1958}
F.~R. Morgenthaler, ``Velocity modulation of electromagnetic waves,'' \emph{IRE Trans. Microw. Theory Tech.}, vol.~6, no.~2, pp. 167--172, Apr. 1958.

\bibitem{mirmoosa2024temporal}
M.~Mostafa, M.~Mirmoosa, M.~Sidorenko, V.~Asadchy, and S.~Tretyakov, ``Temporal interfaces in complex electromagnetic materials: {A}n overview,'' \emph{Opt. Mater. Express}, vol.~14, no.~5, pp. 1103--1127, 2024.

\bibitem{fink2016timereversal}
V.~Bacot, M.~Labousse, A.~Eddi, M.~Fink, and E.~Fort, ``Time reversal and holography with spacetime transformations,'' \emph{Nat. Phys.}, vol.~12, no.~10, pp. 972--977, 2016.

\bibitem{shlivinski2018beyond}
A.~Shlivinski and Y.~Hadad, ``Beyond the {B}ode-{F}ano bound: {W}ideband impedance matching for short pulses using temporal switching of transmission-line parameters,'' \emph{Phys. Rev. Lett.}, vol. 121, no.~20, p. 204301, 2018.

\bibitem{mendoncca2003temporal}
J.~Mendon{\c{c}}a, A.~Martins, and A.~Guerreiro, ``Temporal beam splitter and temporal interference,'' \emph{Phys. Rev. A}, vol.~68, no.~4, p. 043801, 2003.

\bibitem{mendoncca2000quantum}
J.~Mendon{\c{c}}a, A.~Guerreiro, and A.~M. Martins, ``Quantum theory of time refraction,'' \emph{Phys. Rev. A}, vol.~62, no.~3, p. 033805, 2000.

\bibitem{pendry2024air}
J.~B. Pendry, ``Air conditioning for photons,'' \emph{Opt. Mater. Express}, vol.~14, no.~2, pp. 407--413, 2024.

\bibitem{akbarzadeh2018inverse}
A.~Akbarzadeh, N.~Chamanara, and C.~Caloz, ``Inverse prism based on temporal discontinuity and spatial dispersion,'' \emph{Opt. Lett}, vol.~43, no.~14, pp. 3297--3300, 2018.

\bibitem{tien1958parametric}
P.~Tien, ``Parametric amplification and frequency mixing in propagating circuits,'' \emph{J. Appl. Phys.}, vol.~29, no.~9, pp. 1347--1357, 1958.

\bibitem{engheta2020antireflection}
V.~{}Pacheco-Pe{\~n}a and N.~Engheta, ``Antireflection temporal coatings,'' \emph{Optica}, vol.~7, no.~4, pp. 323--331, 2020.

\bibitem{engheta2020aiming}
V.~Pacheco-Pe{\~n}a and N.~Engheta, ``Temporal aiming,'' \emph{Light: Sci. Appl.}, vol.~9, no.~1, p. 129, 2020.

\bibitem{hilbert2009temporal}
S.~A. Hilbert, C.~Uiterwaal, B.~Barwick, H.~Batelaan, and A.~H. Zewail, ``Temporal lenses for attosecond and femtosecond electron pulses,'' \emph{Proc. Natl. Acad. Sci.}, vol. 106, no.~26, pp. 10\,558--10\,563, 2009.

\bibitem{reck2017dirac}
P.~Reck, C.~Gorini, A.~Goussev, V.~Krueckl, M.~Fink, and K.~Richter, ``Dirac quantum time mirror,'' \emph{Physical Review B}, vol.~95, no.~16, p. 165421, 2017.

\bibitem{goldman2014periodically}
N.~Goldman and J.~Dalibard, ``Periodically driven quantum systems: effective {H}amiltonians and engineered gauge fields,'' \emph{Phys. Rev. X}, vol.~4, no.~3, p. 031027, 2014.

\bibitem{dong2024quantum}
Z.~Dong, H.~Li, T.~Wan, Q.~Liang, Z.~Yang, and B.~Yan, ``Quantum time reflection and refraction of ultracold atoms,'' \emph{Nat. Photonics}, vol.~18, no.~1, pp. 68--73, 2024.

\bibitem{ok2024electron}
F.~Ok, A.~Bahrami, and C.~Caloz, ``Electron scattering at a potential temporal step discontinuity,'' \emph{Sci. Rep.}, vol.~14, no.~1, p. 5559, 2024.

\bibitem{bolotovskiui1972superlum}
B.~M. Bolotovski{\u\i} and V.~L. Ginzburg, ``The {V}avilov-{C}erenkov effect and the {D}oppler effect in the motion of sources with superluminal velocity in vacuum,'' \emph{Soviet Phys. Uspekhi}, vol.~15, no.~2, p. 184, 1972.

\bibitem{deck2019uniform}
Z.-L. Deck-L{\'e}ger, N.~Chamanara, M.~Skorobogatiy, M.~G. Silveirinha, and C.~Caloz, ``Uniform-velocity spacetime crystals,'' \emph{Adv. Photonics}, vol.~1, no.~5, p. 056002, 2019.

\bibitem{pendry2022crossing}
J.~B. Pendry, P.~A. Huidobro, M.~Silveirinha, and E.~Galiffi, ``Crossing the light line,'' \emph{Nanophotonics}, vol.~11, no.~1, pp. 161--167, 2022.

\bibitem{granatstein1976realization}
V.~Granatstein, P.~Sprangle, R.~Parker, J.~Pasour, M.~Herndon, S.~Schlesinger, and J.~Seftor, ``Realization of a relativistic mirror: {E}lectromagnetic backscattering from the front of a magnetized relativistic electron beam,'' \emph{Phys. Rev. A}, vol.~14, no.~3, p. 1194, 1976.

\bibitem{lampe1978interaction}
M.~Lampe, E.~Ott, and J.~H. Walker, ``Interaction of electromagnetic waves with a moving ionization front,'' \emph{Phys. Fluids}, vol.~21, no.~1, pp. 42--54, 1978.

\bibitem{hadad2024space}
Y.~Hadad and D.~Sounas, ``Space-time modulated loaded-wire metagratings for magnetless nonreciprocity and near-complete frequency conversion,'' \emph{Opt. Mater. Express}, vol.~14, no.~5, pp. 1295--1308, 2024.

\bibitem{taravati2017nonreciprocal}
S.~Taravati, N.~Chamanara, and C.~Caloz, ``Nonreciprocal electromagnetic scattering from a periodically space-time modulated slab and application to a quasisonic isolator,'' \emph{Phys. Rev. B}, vol.~96, no.~16, p. 165144, 2017.

\bibitem{estep2014magnetic}
N.~A. Estep, D.~L. Sounas, J.~Soric, and A.~Al\`{u}, ``Magnetic-free non-reciprocity and isolation based on parametrically modulated coupled-resonator loops,'' \emph{Nat. Phys.}, vol.~10, no.~12, pp. 923--927, 2014.

\bibitem{Deck_PRB_2018}
Z.-L. Deck-L\'{e}ger, A.~Akbarzadeh, and C.~Caloz, ``Wave deflection and shifted refocusing in a medium modulated by a superluminal rectangular pulse,'' \emph{Phys. Rev. B}, vol.~97, no.~10, pp. 104\,305--1:7, Mar. 2018.

\bibitem{taravati2019generalized}
S.~Taravati and G.~V. Eleftheriades, ``Generalized space-time-periodic diffraction gratings: {T}heory and applications,'' \emph{Phys. Rev. Appl.}, vol.~12, no.~2, p. 024026, 2019.

\bibitem{cassedy1963dispersion}
E.~S. Cassedy and A.~A. Oliner, ``Dispersion relations in time-space periodic media: {P}art {I}—{S}table interactions,'' \emph{Proc. IEEE}, vol.~51, no.~10, pp. 1342--1359, 1963.

\bibitem{cassedy1967dispersion}
E.~S. Cassedy, ``Dispersion relations in time-space periodic media: {P}art {II}—{U}nstable interactions,'' \emph{Proc. IEEE}, vol.~55, no.~7, pp. 1154--1168, 1967.

\bibitem{chamanara2017optical}
N.~Chamanara, S.~Taravati, Z.-L. Deck-L{\'e}ger, and C.~Caloz, ``Optical isolation based on space-time engineered asymmetric photonic band gaps,'' \emph{Phys. Rev. B}, vol.~96, no.~15, p. 155409, 2017.

\bibitem{huidobro2019fresnel}
P.~A. Huidobro, E.~Galiffi, S.~Guenneau, R.~V. Craster, and J.~B. Pendry, ``Fresnel drag in space--time-modulated metamaterials,'' \emph{Proc. Natl. Acad. Sci.}, vol. 116, no.~50, pp. 24\,943--24\,948, 2019.

\bibitem{huidobro2021homogenization}
P.~A. Huidobro, M.~G. Silveirinha, E.~Galiffi, and J.~Pendry, ``Homogenization theory of space-time metamaterials,'' \emph{Phys. Rev. Appl.}, vol.~16, no.~1, pp. 014\,044:1--13, 2021.

\bibitem{Lurie_Springer_2007}
K.~A. Lurie, \emph{An {I}ntroduction to the {M}athematical {T}heory of {D}ynamic {M}aterials}.\hskip 1em plus 0.5em minus 0.4em\relax Springer, 2007.

\bibitem{joannopoulos2003shockwave}
E.~J. Reed, M.~Solja{\v{c}}i{\'c}, and J.~D. Joannopoulos, ``Color of shock waves in photonic crystals,'' \emph{Phys. Rev. Lett.}, vol.~90, no.~20, p. 203904, 2003.

\bibitem{joannopoulus2022accelerating}
J.~Sloan, N.~Rivera, J.~D. Joannopoulos, and M.~Solja{\v{c}}i{\'c}, ``Controlling two-photon emission from superluminal and accelerating index perturbations,'' \emph{Nat. Phys.}, vol.~18, no.~1, pp. 67--74, 2022.

\bibitem{bahrami2023generalized}
A.~Bahrami, Z.-L. Deck-L{\'e}ger, Z.~Li, and C.~Caloz, ``A generalized {FDTD} scheme for moving electromagnetic structures with arbitrary space-time configurations,'' \emph{IEEE Trans. Antennas Propag.}, vol.~72, no.~2, pp. 1721--1734, 2024.

\bibitem{bahrami2023bending}
A.~Bahrami and C.~Caloz, ``Electrodynamics of accelerated space-time engineered-modulation metamaterials,'' in \emph{2023 Metamaterials}.\hskip 1em plus 0.5em minus 0.4em\relax IEEE, 2023, pp. X--031.

\bibitem{bahrami2023electrodynamics}
A.~Bahrami, Z.-L. Deck-L{\'e}ger, and C.~Caloz, ``Electrodynamics of accelerated-modulation space-time metamaterials,'' \emph{Phys. Rev. Appl.}, vol.~19, no.~5, p. 054044, 2023.

\bibitem{caloz2019spacetime1}
C.~Caloz and Z.-L. Deck-L{\'e}ger, ``Spacetime metamaterials—{P}art {I}: {G}eneral concepts,'' \emph{IEEE Trans. Antennas Propag.}, vol.~68, no.~3, pp. 1569--1582, 2019.

\bibitem{caloz2019spacetime2}
C.~{}Caloz and Z.-L. Deck-L{\'e}ger, ``Spacetime metamaterials—{P}art {II}: Theory and applications,'' \emph{IEEE Trans. Antennas Propag.}, vol.~68, no.~3, pp. 1583--1598, 2019.

\bibitem{bellman1966slab}
R.~Bellman, R.~Kalaba, and S.~Ueno, ``Invariant imbedding and scattering of light in a one-dimensional medium with a moving boundary,'' \emph{J. Math. Anal. Appl.}, vol.~15, no.~2, pp. 171--182, 1966.

\bibitem{tsai1967wave}
C.~Tsai and B.~Auld, ``Wave interactions with moving boundaries,'' \emph{J. Appl. Phys.}, vol.~38, no.~5, pp. 2106--2115, 1967.

\bibitem{ishimaru2017electromagnetic}
A.~Ishimaru, \emph{Electromagnetic Wave Propagation, Radiation, and Scattering: From Fundamentals to Applications}.\hskip 1em plus 0.5em minus 0.4em\relax John Wiley \& Sons, 2017.

\bibitem{ostrovskiicorrect1967}
L.~A. Ostrovskii and B.~A. Solomin, ``Correct formulation of the problem of wave interaction with a moving parameter jump,'' \emph{Radiophys. Quantum Electron.}, vol.~10, no.~8, pp. 666--668, 1967.

\bibitem{deck2019interluminal}
Z.-L. Deck-L{\'e}ger and C.~Caloz, ``Scattering at interluminal interface,'' in \emph{2019 AP-S}.\hskip 1em plus 0.5em minus 0.4em\relax IEEE, 2019, pp. 367--368.

\bibitem{supp_mat}
{S}ee Supplemental Material at [URL will be inserted by publisher] for the detailed derivation of the equations in the text.

\bibitem{van2012relativity}
J.~Van~Bladel, \emph{Relativity and {E}ngineering}.\hskip 1em plus 0.5em minus 0.4em\relax Springer Science \& Business Media, 2012, vol.~15.

\end{thebibliography}
}
\end{document}